\documentclass[12pt]{article}
\input epsf.sty
\def\be{\begin{equation}} \def\ee{\end{equation}}
\def\bea{\begin{eqnarray}} \def\eea{\end{eqnarray}} \def\ba{\begin{array}}
\def\ea{\end{array}} \def\ben{\begin{enumerate}} \def\een{\end{enumerate}}
  
\def\lll{\label}
\newcommand{\eqn}[1]{(\ref{#1})}

\newcommand{\prd}[3]{Phys. Rev. {\bf D#1} ({#2}) {#3}}
\newcommand{\hepth}[1]{{\tt hep-th/{#1}}}

\def\k{\kappa}

\def\ov{\over}
\def\br{\nonumber\\}

\begin{document}
{}~
\hfill\vbox{\hbox{hep-th/0508101} \hbox{SINP/TNP/05-20}}\break

\vskip .5cm
\centerline{\large \bf
(A)symmetric  Tachyon  Rolling in de Sitter Spacetime: }
\centerline{\large \bf
A universe devoid of Planck density }

\vskip .5cm

\vspace*{.5cm}

\centerline{ \sc Harvendra Singh}

\vspace*{.5cm}

\centerline{ \it Theory Division, Saha Institute of Nuclear Physics,} 
\centerline{ \it 1/AF Bidhannagar, Kolkata 700064, INDIA}

\vskip.5cm
\centerline{[E-mail: h.singh@saha.ac.in]}

\vskip1cm
\centerline{\bf Abstract} \bigskip

In a recent paper we studied rolling tachyon
flat FRW cosmologies, but those admitting only time-reversal asymmetric 
boundary conditions. The time-reversal symmetric 
cosmologies have been studied by Sen previously.  
We show explicitly here that through appropriate choice of initial 
conditions, the time evolution of the
Hubble parameter in
these two types of solutions can be made completely identical for $t>0$, 
except near $t=0$. The rolling tachyon solution also gives rise to 
 necessary inflation. 
We find that universe does start as a string size object (with string 
scale $10^{15} GeV$)  
with a string mass density $\rho_s\simeq 10^{78} gm. cm^{-3}$ and not with 
Planck density.

\vfill \eject

\baselineskip=16.2pt

\section{Introduction}
Recently, the phenomenon of tachyon condensation \cite{sen1,sen2}  
has been a subject of much attention in string theory and as well as in 
cosmological applications \cite{sen3,linde3}. The  tachyon 
field appears as 
instability in D$p$-$\bar {\rm D} p$-branes or non-BPS D$p$-brane 
systems 
in type II string theory.\footnote{There are closed string 
tachyons in bosonic string theory as well; see
\cite{yang,yang1} for recent developments.} The 
low energy dynamics of the open string tachyons living on the world-volume 
of unstable D3-brane is governed by
the DBI  action
\cite{sen0,sen1,sen2}
\be\lll{tachyact}
-\int d^4x V(T)\sqrt{-{\rm det} (g_{\mu\nu}+\alpha '\partial_\mu 
T\partial_\nu 
T)} \ee
where tachyon $T$ appears explicitly as a world volume scalar 
field, 
$g_{\mu\nu}$ represents the pull-back of the spacetime metric. $V(T)$ 
is a positive definite tachyon potential which has  a maximum while
it vanishes as $T\to\infty$. The Regge 
slop parameter
$\alpha'=(l_s)^2=(M_s)^{-2}$ where $l_s$ is the string length   
and $M_s$ is the string mass. 
In a flat spacetime background the equation 
of motion 
for a purely time-dependent tachyon field assumes following unique form:
\be\lll{a1}
{V(T)\over\sqrt{1-\alpha '\dot T^2}}\propto\rho.
\ee
It means that the tachyon field behaves like a fluid of 
constant positive energy 
density, $\rho$, and of negative pressure, $P=-V\sqrt{1-\alpha '\dot 
T^2}$. 
Indeed, if at the top of the potential 
 $\dot T=0$, we have got an useful example of the cosmological constant
 for which
$\rho=-P=constant$. 
During  
the time evolution tachyon field rolls down the potential and reaches its 
true vacuum state given by $T=\infty$ where $V(\infty)=0$, while keeping 
$\rho$ 
 constant through out. The pressure also vanishes in the tachyon 
vacuum. Thus the end state behaves like a pressureless matter (dust). 
When the tachyon system is 
coupled to  nontrivial closed string background
this result does get significantly modified and new type of 
cosmological solutions can be obtained. In recent times   
the application of open string tachyons in inflationary 
cosmological models has gained favourable
attention; see the references
\cite{anup,gibbon,faityt,cgjs,linde,sami,irina,peet,roy,odintsov, 
kklt2,garousi,sen3,linde3,linde2,hs5,sinha}.
\footnote{
A wide class of useful references on this topic can also be found in 
\cite{sen3,linde3,paddy}.}

In the previous paper \cite{hs5} we studied a very specific model 
in which tachyon action was coupled to 
de Sitter background. Especially we studied 
the flat homogeneous FRW cosmological solutions with time-reversal 
non-symmetric initial conditions. The 
solutions found there were completely {\it non-singular} and  described 
inflationary situations just like after the big-bang.
In this paper we would like to study the 
time-reversal symmetric model of Sen \cite{sen3} and compare it to the 
time-reversal asymmetric situations in \cite{hs5}. 
We find surprising 
similarities in the two cases.
In particular we find it hard to distinguish, from 
the point in time where we are 
today, whether our universe has been time-reversal 
symmetric or otherwise.  
We do  also provide a detailed analysis of 
the inflation which has a greater appeal 
for  phenomenological 
applications. The crucial point in our analysis is that we 
are specially considering a large 
number density of non-BPS branes over the compactified space. The large 
density of branes helps us in getting suitably high value of the 
Hubble constant during inflation.  The previous attempts on 
inflationary tachyonic models 
however study the effects of spacetime warping on the tachyon 
potential; see \cite{sinha} and references there in.

The paper is organised as follows. In section-2  we review and discuss
the  FRW cosmological models obtained by  
coupling bulk de Sitter supergravity to the tachyon action. We study  
the solutions both with and 
without time-reversal symmetric 
conditions. In section-3 we provide 
complete numerical analysis of
 simple inflationary models.      
 We  estimate the  number of e-foldings during the  
inflation in realistic situations and fix the values of various parameters 
so that the results are close to the observed values.  
In section-4  we  discuss  the slow-roll limits coming 
from cosmology. 
We summarise the main results in 
section-5.

\section{(A)symmetric de Sitter cosmology with tachyon field}
 
We review the model considered by Sen \cite{sen3}  based on tachyon
field theory coupled to background gravity and nonvanishing positive
cosmological constant, $\Lambda$. 
It is significant in these models that  compactification of 
string theory to four-dimensions with a positive cosmological constant, of 
KKLT type \cite{kklt}, 
is  achieved. All 
the moduli including the volume modulus are taken to be apriori fixed. The 
model also has an unstable D3-brane  
 which extends along the three large spatial directions. 
The four-dimensional effective action is taken  as \cite{sen3}
\be\lll{1a}
S=\int d^4x \bigg[ {1\over 16\pi G}\sqrt{-g}( R 
-2\Lambda)-V(T)\sqrt{-{\rm det} (g_{\mu\nu}+\alpha'\partial_\mu 
T\partial_\nu T)}
\bigg]
\ee
where Newton's constant $G$ is related to   
four-dimensional Planck mass $M_p$ as $8\pi G= M_p^{-2}$.
We shall consider that at the top of the potential $V(T_0)=V_0$,  
while $V(\infty)=0$ in the 
tachyon vacuum. These are the two 
well known properties of the tachyon potential, e.g.
$V(T)=V_0/{\rm cosh}(T/\sqrt{2})$ in superstring theory \cite{sen2}.
\footnote{Note that we shall take $V_0\sim {N T_3}$ for this 
potential, where $N$ is the 
number of 3-branes. We are considering more than one 
unstable 3-brane or equivalently a fat 3-brane. Here $T_3= 
{\sqrt{2}M_s^4\over 
(2\pi)^3g_s} $ is the
non-BPS D3-brane tension.} 

The recent cosmological observations tell us
that our universe is isotropic and homogeneous at large 
scales and also appears to be spatially flat \cite{wmap}. These 
measurements 
also reveal that the present phase of the expansion is such that
the vacuum energy (the cosmological constant) contributes nearly $70\%$ to 
the total energy density in the universe ($\Omega_{total}=1$).
Motivated by these phenomenological inputs we 
shall always assume $G V_0\gg \Lambda $. 
We look for purely time-dependent  solutions of the 
equations of motion and shall take the metric ansatz to be a
generic Friedman-Robertson-Walker spacetime 
\be\label{1a1}
ds^2=-dt^2 +a(t)^2\left({dr^2\over 1-k r^2}+r^2(d\theta^2+sin^2\theta 
d\phi^2)\right)
\ee
where $k$ takes values $1,~0$ or $-1$ depending upon whether the spatial
geometry is
$S^3,~R^3$ or $H^3$ respectively. We also take  tachyon 
to be purely time-dependent $T=T(t)$. 

With the above ans\"atze  
the field equations derived from the above action 
can be written as; the Tachyon equation:
\be\lll{teqn}
\ddot T=-(1-\alpha'\dot T^2)\left(M_s^2{V'\over V}+3\dot T {\dot a\over 
a}\right) 
\ee
the Friedman equation:
\be\lll{feqn}
{\ddot a \over a}= {\Lambda\ov 3}+{8\pi G\ov
3}\bigg[{V(T)\over\sqrt{1-\alpha'\dot T^2}}(1 -{3\ov2} \alpha'\dot 
T^2)\bigg] 
\ee
and the Raychaudhuri equation:
\be\lll{reqn}
\left({\dot a\over a}\right)^2=-{k\over a^2}+ {\Lambda\ov 3}+{8\pi G\ov
3}{V(T)\over\sqrt{1-\alpha'\dot T^2}}.
\ee
It can be seen that the equation \eqn{teqn} 
follows simply by taking the time derivative of the eq. \eqn{reqn}. 
Therefore equations \eqn{feqn} and \eqn{reqn} contain all information 
about tachyon rolling in this model, making \eqn{teqn} redundant. 
Let us introduce a quantity $H(t)\equiv\dot a(t)/a(t)$  known as 
Hubble parameter at any given time $t$. Then equations
\eqn{feqn} and \eqn{reqn}
immediately give us 
\be\lll{keqn}
\dot H={\k\over a^2}-{3\over 2} {8\pi G\ov
3}{V(T)\over\sqrt{1-\alpha'\dot T^2}} \alpha'\dot T^2
\ee
This is the key equation for the evolution of the Hubble parameter.

It can be seen that there is a time-reversal symmetry of these field 
 equations if the fields (solutions) have the following properties 
\cite{sen3}
\be
~~T(t)= T(-t),~~H(t)= -H(-t),~~a(t)= a(-t).
\ee 
A generic time-reversal 
symmetric solution with boundary conditions \cite{sen3} 
\be\lll{sym}
 ~ T(0)=T_0,~\dot 
a(0)=0,~a(0)=a_0,~\dot T=0
\ee
at $t=0$ necessarily requires $k=1$; 
that is, the  spatial FRW geometry has to be $S^3$. However, for 
the time-asymmetric 
initial conditions \cite{hs5}
\be\lll{asym}
 ~ T(0)=T_0,~H(0)=H_0,~\dot T(0)=0
\ee
at $t=0$ we can set $k=0$, which corresponds to a flat spatial geometry. 

Note that the second term on the right hand 
side of \eqn{keqn} is always negative definite, 
while $\alpha'\dot T^2$ can only
vary between 0 and 1 monotonically. 
We clearly see from \eqn{keqn}, keeping in mind the conditions \eqn{sym}, 
that initially
${\dot H}>0$ where $k\over a^2$ term is dominant while at the 
later stage the tachyon terms take over
making ${\dot H}<0$ {\it always}.
While 
specially for $k=0$ (time-asymmetric) case ${dH\over dt}<0$ always 
\cite{hs5}.
Thus we also infer from here that the late time 
evolutions are the similar in the two cases, that is, 
 at the later stage in
tachyon evolution, 
Hubble constant always decreases with time or at most becomes
constant (when $\Lambda\ne0$), but it cannot increase in  time. 

\subsection{dS $\leftrightarrow$ dS Vacuum Interpolation}
There are mainly two exact de Sitter solutions of the equations of 
motion in the 
last section with a constant tachyon field. The de Sitter solution with 
a bigger value of cosmological constant $\Lambda_0={{\Lambda+ 8\pi  G 
V_0}}$ is given by \cite{sen3}
\be\lll{sym1}
 T=0,~~\dot T=0,~
a(t)=a_0{\rm cosh}(\sqrt{\Lambda_0/3}~t),~~ a_0^2=3/\Lambda_0
\ee
and the other de Sitter solution is
\be\lll{asym1}
 T=\infty,~~\dot T={1\over\sqrt{\alpha'}},~
a(t)=a_0{\rm cosh}(\sqrt{\Lambda/3}~t),~~ a_0^2=3/\Lambda.
\ee
Also, there will be  other rolling tachyon solutions of the field 
equations which 
interpolate between these two de Sitter spaces.
(Specially, if we set $\Lambda=0$, then the interpolation will be between 
dS 
and 
Minkowski flat spacetime.) This can be immediately seen over here. 
With the initial conditions \eqn{sym},  in the neighborhood 
of $t=0$ where $\dot T\simeq0$, the field equations simply reduce to
\bea\lll{eqa1}
&&H^2=-{\k\over a^2}+ {\Lambda\ov 3}+{8\pi G\ov
3}{V_0} + {  O} (\alpha'\dot T^2) \br
&& \dot H= 
{\k\over a^2}- {  O} (\alpha'\dot T^2) 
\eea
These are precisely the equations of a de Sitter spacetime with 
effective cosmological constant 
$ \Lambda_0 $. Hence the universe near the 
top of the potential inflates
very fast as a  de Sitter space.
 This {\it pure} de Sitter phase lasts  for a 
very short time until  $\dot T^2$ terms in \eqn{eqa1} become sizable.
We will also see from numerical analysis that in this short interval 
the Hubble parameter reaches its maximum value 
given by $$H_m=\sqrt{\Lambda_0\over3}$$ starting from the zero value. 
As $\dot T$ grows in time, 
from eq. \eqn{keqn}, at some instant, $\dot 
H(t)\simeq0$. While 
 ever after that we will have decelerating (matter dominated) phase $\dot 
H<0$. 
In the far future (as $t\to\infty$) 
where   
$T\simeq\infty$ and $\alpha'\dot T^2\simeq 1$, the field equations  
simplify to
\bea\lll{eqa2}
&&H^2=-{\k\over a^2}+ {\Lambda\ov 3},~~ \dot H= 
{\k\over a^2} 
\eea
which is the de Sitter universe with a cosmological constant $\Lambda$. 
Thus the models with both the boundary conditions \eqn{sym} or 
\eqn{asym} will interpolate between two de Sitter phases with  
cosmological constants $\Lambda_0$ and $\Lambda$ respectively. 
The value of $\Lambda$ could as well be chosen to be the present 
value of
cosmological constant in our universe, being
$\sim 10^{-122} M_p^2$. To note, in the time-reversal symmetric case
the Hubble parameter must change sign across $t=0$. 
Therefore universe must start with a (contracting) de Sitter phase at 
$t=-\infty$ and should end 
in the (expanding) de Sitter universe at $t=\infty$ separated by an 
intermediate de 
Sitter phase at $t=0$ \cite{sen3}.

\section{Numerical analysis of the tachyon rolling}
\subsection{Basic analysis}
An interpolating solution of the tachyon-gravity system
as described in the last section will not be easy to obtain analytically.  
Therefore
our main aim  is to  solve  the
tachyon-gravity system numerically. For simplification
 we set $\Lambda=0$, because for $\Lambda\ll 
G V_0$ the basic results 
will be similar except they will differ only at the late time. 
The equations \eqn{reqn} and  \eqn{keqn} can  be assembled in
the following  form

\bea\lll{aeqn1}
{\dot H -{k\over a^2}\over H^2+{k\over a^2}}=-{3\over 2}\alpha' 
\dot T^2 \ , ~~
{8\pi G\ov
3}{V(T)\over\sqrt{1-\alpha'\dot T^2}}-{k\over a^2}=H^2 .
\eea

First
we are interested in the initial conditions \eqn{asym} which represent 
the time asymmetric situation ($k=0$) \cite{hs5}. We take $ T(0)=0,~ 
H(0)=H_0$
such that at the top of the potential $\dot T(0)=0$. Then eqs.\eqn{aeqn1} 
imply $\dot H(0)=0$ with ${8\pi G\ov
3}V(0)=H_0^2$. Classically the system with these initial 
conditions will not evolve in time, however a 
small fluctuation will dislocate the configuration from the top 
and the system will eventually start evolving. We shall take initial 
conditions in 
our numerical analysis in conformity with this fact where we are slightly 
away 
from the top position. We consider 
$V(T)=V_0/{\rm cosh}({T\over\sqrt{2}})$ in these calculations for both 
symmetric and asymmetric cases.  
In our units $t$ is measured in $M_p^{-1}$ and 
the Hubble constant $H$ is measured in $M_p$. Accordingly the string 
tension parameter
$\alpha'$ 
which multiplies $\dot T^2$ has to be fixed. From string compactification
\be\lll{MsMp}
M_s^2={g_s^2\over v_0}M_p^2 ,
\ee
 with the volume parameter
 $v_0\equiv ({R \over l_s})^6{1\over\pi}$ ($R$ 
being 
radius of compactification).  Since in this subsection, we are interested 
in the qualitative understanding of the tachyon rolling in symmetric and 
asymmetric cases, 
we just set ${g_s^2\over v_0}=1$, so that $M_s=M_p$. Notice that    
${g_s^2\over v_0}=1$ corresponds to small volume compactification and so 
the results will be sensitive to stringy corrections. Therefore, for 
 realistic 
scenarios we shall always consider large volume compactification in the 
next subsection.  

\begin{figure}[!ht]
\leavevmode
\begin{center}
\epsfysize=5cm
\epsfbox{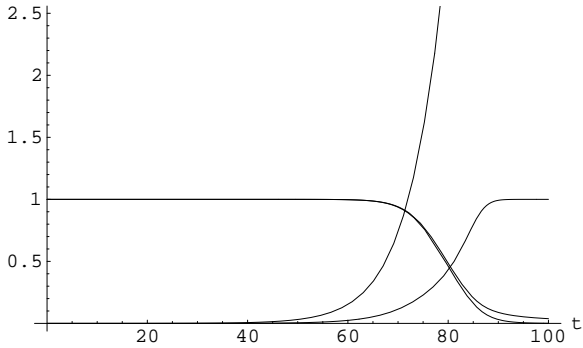}
\end{center}
\caption{\it Plots are for the time evolution of $T,~\dot T,~H$ and 
$\sqrt{8\pi G V / 3} $ presented in clockwise manner, with initial 
values  $H(0) = 1$, $T(0) =10^{-5}$. 
The value of $N_e$, which is the area under $H(t)$ curve in the 
plateau region, could easily 
be estimated to be close to 60. 
 $H$ vanishes at late time but $V(T)$ 
vanishes faster than $H$.} \label{fig1}
\end{figure}

\begin{figure}[!ht]
\leavevmode
\begin{center}
\epsfysize=5cm
\epsfbox{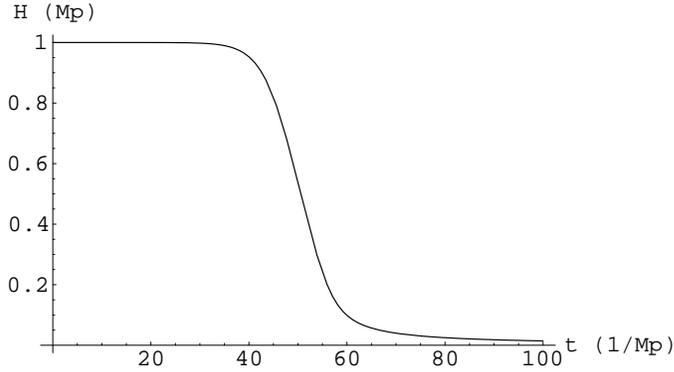}
\end{center}
\caption{ \it The graph 
is for $H$  with initial values 
$H(0)=1 (M_p)$ and  $T(0) =10^{-3}$. 
The value of $N_e$ can
be estimated to be about 50 in this case. $H$ vanishes only at late time 
but earlier than in the previous graph.
}\label{fig2}
\end{figure}

\begin{figure}[!ht]
\leavevmode
\begin{center}
\epsfysize=5cm
\epsfbox{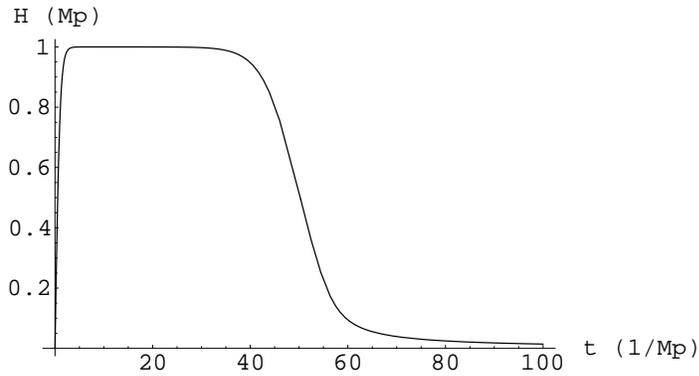}
\end{center}
\caption{\it For comparison the 
plot is  for $H(t)$ 
with time-reversal symmetric initial values 
$H(0)=0,~T(0) =10^{-3}$ and ${8\pi G 
V(0) \over 3}=1$. 
The value of $N_e$ is
estimated to be 50 in this case. Note $H$ 
starts from initial value zero at $t=0$. } \label{fig3}
\end{figure}

The numerical results for two different time-asymmetric initial conditions
\eqn{asym} are plotted in the 
figures (\ref{fig1}) and 
(\ref{fig2}). Specially
the time evolution of the various quantities $T,~\dot T,~H$ and 
$\sqrt{8\pi G V(T) / 3} $ could be found in the figure (\ref{fig1}). The 
plots of 
$H$ and $\sqrt{8\pi G V \over 3} $ do overlap with each other in the 
plateau
region. This plateau region we will characterise as the {\it slow-roll} 
inflationary period. Note that
both $H$ and $V$ vanish at the late time, but $V(T)$ 
vanishes faster than $H$. This fact led us to propose the bound 
\footnote{
The number of e-foldings through which universe expands during the 
slow-roll inflation is defined as
$N_e={\rm ln}{a(t_f)\over a(t_i)}\equiv\int_{t_i}^{t_f} H(t)dt$, where 
$t_i$ is the time when inflation starts and $t_f$ is the time when 
inflation ends. The estimate is that 
the universe went through  at least $60$ e-foldings.}
$$N_e \ge \int_{t_i}^{t_f} \sqrt{8\pi G V \over 3}dt$$ 
in the previous work 
\cite{hs5}.
From fig.(\ref{fig1}), it 
is found that taking the initial values $H(0)\simeq 1 M_p$ and the 
$T(0)\simeq 
10^{-5}$ provide us with the number of e-folds during inflation close to 
$60$.
The duration of inflation is rather very short and is about $60 
(M_p)^{-1}$ only.
But it can be increased by taking smaller values of $H(0)$ and 
$T(0)$, see \cite{hs5}. At least 
from figures (\ref{fig1}) and (\ref{fig2}) it can be deduced  that lower  
the initial value  
$T(0)$  longer is the duration of inflation. 

Now we come to {\it time-reversal symmetric} evolution. We 
take the 
boundary conditions as in \eqn{sym} along with $k=1$ in \eqn{aeqn1}. 
The time evolution of 
 $ H(t)$  is plotted in 
fig.\ref{fig3},  keeping the initial values $T(0)$ and ${8\pi G 
V(0) \over 3}$ same as for fig.\ref{fig2}. We can see 
from the figure (\ref{fig3}) that near $t=0$ 
the Hubble 
parameter grows very-very fast starting from a initial zero value and in 
no time it 
reaches its maximum value of 1. After that it remains constant 
for a while,
this constant $H$ reason where $H$ is also maximum will be characterised 
as the slow-roll inflation. 
Later the matter 
dominated (decelerating) phase takes over which will terminate in a  
de Sitter 
phase as $t\to\infty$. Due to time-reversal symmetry, the plot must be 
extended to negative $t$ axis also, but we have not plotted it.  
However, the inflation lasts for a rather short
interval of 
$50 M_p^{-1}$ only. Moreover, the value of the Hubble parameter at the end 
of inflation is of the order of $ 1 M_p$ which is much larger value than 
the COBE 
bound $10^{-5} M_p$ from the fluctuations. So these $H_0\sim O(M_p)$ 
models are not good and require a change in the various initial values of 
the parameters and fields. 
	
{\it Thus,
quite significantly, comparing figures (\ref{fig2}) and (\ref{fig3}) 
 which have same initial conditions except for the value  $\dot a(0)$, 
we learn that the 
time-symmetric and
asymmetric evolutions of the Hubble parameter differ in picture only in 
the neighborhood of $t\simeq 
0$, but not at  later (positive) time. That is, by observing the 
universe at late 
times we may not be able to distinguish between the two situations.}

\subsection{A Near Phenomenological Choice of Parameters}
The analysis of the last section looks very promising in that the tachyon 
rolling does seem to provide, albeit qualitatively, the picture of 
inflationary  evolution of our 
universe. In order to get a realistic picture we need to pick those 
values of 
parameters $g_s,v_0,N$ which  would restrict 
the size of perturbations 
$\delta_H$ (or density 
fluctuations) to less than $10^{-5}M_p$ towards the end of 
slow-roll inflation. This will mostly
depend on  the cosmological constant $\Lambda_0$ of the de Sitter phase 
near $t=0$
which controls the value of the Hubble parameter during inflation (the 
plateau region in the figures). The value of  $\Lambda_0$ depends upon the 
height of  
tachyon potential
through $ \Lambda_0=\Lambda+8\pi G V_0$. From the stringy origin of the 
tachyon potential, 
$V$, we find 
 that the maximum value of Hubble parameter is (as $\Lambda\ll 8\pi G 
V_0$) 
\be\lll{ae1}
(H_m)^2=\Lambda_0/ 3\simeq {8\pi G V_0\over3} \approx 
{N\over g_s}({g_s^2\over v_0})^2 M_p^2.\ee
where $N$ counts the number of  3-branes.\footnote{Note we have 
dropped a factor of 
${\sqrt{2}\over3 (2\pi)^3}$  on the RHS of the 
above equation. Note, it can be easily absorbed in the parameter $N$  
with out affecting the whole analysis.} To bring stringy corrections under
control, we 
shall consider  large volume 
compactification, $v_0>1$, with generic weak string 
coupling, $g_s<1$. Using eq.\eqn{MsMp}, the above equation \eqn{ae1} can 
also be written as
\be\label{ae2}
H_m\sim \sqrt{{N g_s\over v_0}} M_s\ .
\ee
Thus, generically we will have a situation $H_m<<M_s<<M_p$
whenever $g_S<1$ and $v_0>1$.
However, from our experience in the last section, for suitable 
inflation, cases with $H_m {\simeq} M_s$ are more favourable. From 
\eqn{ae2}, this can be achieved only if we
consider suitably large number of the branes so that
 ${N g_s\over v_0} { \sim} 1$. In the following it will be our primary 
assumption.

Now,  particularly if we take $g_s=0.1$,
${v_0}= 10^{6}$, and $N/v_0 = 10$, then the  
string scale  becomes $M_s= 10^{-4}M_p$ and 
\be\lll{sit1}
H_m {\sim} M_s \simeq
10^{-4} M_p\ .\ee
While if we take 
$g_s=0.1$ and
${v_0}= 10^{6}$, along with $N/v_0 \sim 1$, the 
string scale is still  $M_s= 10^{-4}M_p$ but now 
\be\lll{sit2}
H_m \sim 0.316 M_s\ .
\ee
Both of the situations \eqn{sit1} and \eqn{sit2} appear to be close to 
each other, but as we will see in next section that $H_m<M_s$ is less 
favoured for cosmological applications as this tend to spoil the 
slow-roll. 
So we shall take $H_m{\sim} M_s\sim 10^{-4} M_p$, i.e. \eqn{sit1}, for the 
model which we 
are going to discuss in the rest of the paper. 
Also, keeping
$H_m {\sim} M_s $ the closed string production rate during the 
tachyon decay will be rather suppressed (dilute approximation) compared to 
the situation when 
$H_m\gg M_s$. This may be important as  we are assuming that classical 
action \eqn{1a} remains valid through out the time evolution. 
Notice, however, 
taking 
$H_m {\sim} M_s $ also makes 
the temperature of the initial de Sitter phase comparable to the Hagedorn 
temperature of the strings.\footnote{ We are thankful to the referee for 
suggesting 
this important point. Nevertheless, we are assuming that the classical 
analysis holds good so long as $H\sim M_s$.}    

\noindent{\it \large The Model:}

\begin{figure}[!ht]
\leavevmode
\begin{center}
\epsfysize=5cm
\epsfbox{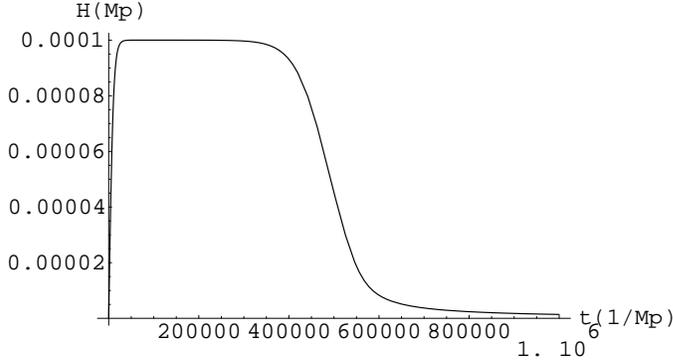}
\end{center}
\caption{ \it The 
plot is for $H(t)$ 
with time-reversal symmetric initial values 
$H(0)=0,~ T(0)=10^{-3}$ with ${8\pi G 
V(0) \over 3}=10^{-8}  $. 
The value of $N_e$ could easily 
be estimated to be 50 in this case. } \label{fig5}
\end{figure}

Based on above approach, in the 
figure (\ref{fig5}) we 
have  plotted an interpolating (numerical) solution 
of the eqs.\eqn{aeqn1} by taking
\be\label{ic1}
g_s=0.1, ~M_s=10^{-4}M_p,~~ 8\pi G V_0=3\times 10^{-8} M_p^2 
\ee
and  the time-symmetric initial conditions ($k=1$) at $t=0$ as
\be\label{ic2}
T(0)=T_0=10^{-3},~ \dot a(0)=0, ~a(0)=\sqrt{3/ (8\pi G V_0)} \ .
\ee
This is just one of the suitable choices. One can take slightly 
different values of these quantities. For example, if we take 
$T(0)\sim .01$ keeping everything else same, the duration of inflation 
will be slightly
reduced  compared to the fig.\ref{fig5}. 
We will discuss more on this in  section-4.
For practical reasons we have set 
$\Lambda=0$ while numerically solving the equations. Also if we
keep  $\Lambda\approx 10^{-122}M_p^2$ in these 
equations it will not alter the major conclusions. 

\noindent{\it \large Critical number density:}

It is important to emphasize that above construction heavily depends upon 
the proposal that we can have a large number of  3-branes 
such that $N/v_0\equiv N({l_s\over R})^6\sim O(1)$. The 
quantity $N/v_0=1$ does quantify a critical 
number density for us. For example, a large compactification radius of 
$R=10\, l_s$  will imply 
$N\sim10^6$. Thus if compactification volume is large so is the number of 
branes, such that {\it there is one 3-brane for each  
 six-volume block (of string-size)
 of the compact six-manifold.} 
If this flexibility is not there then we will not have the construction 
we discussed so far.

\subsection{Inflation and growth of scales }

It will be worth while to know, how the scales grow during the 
inflation and the rest of the expansion in the time symmetric model.
The work is simplified if we divide the whole time evolution from $t=0$ to 
$t=t_{today}$ into definite intervals during which Hubble parameter grows 
differently. Based on the graph in fig.\ref{fig5} the evolution can be 
divided into {\it four} distinct phases of expansion. We shall use the 
values in eqs. \eqn{ic1}, \eqn{ic2} and fig.\ref{fig5} for the estimate 
of various quantities in this section.

\noindent{\it Phase-1 (pure de Sitter)}

During this phase $\dot H>0$, universe grows with an accelerated expansion
from $t=0$ to the time $t_i$ where Hubble parameter reaches its maximum 
value $H_m=\sqrt{\Lambda_0/ 3}$. This is a pure de Sitter phase during 
which scale 
factor grows as $a(t)=a_0 {\rm cosh}(H_m t)$. Thus the expansion lasts 
typically
for the period $\sim (H_m)^{-1}$. For our choice of $H_m\approx 10^{-4} 
M_p$ this period of de Sitter expansion is calculated to be around $10^{4} 
(M_p)^{-1}$. This is quite evident from the fig.\ref{fig5} also. {\it The 
size 
of the universe grows roughly by 10 times of the initial size}. The 
initial size at 
$t=0$ is, however,  fixed by $a_0=\sqrt{3/\Lambda_0}$, which is of the 
string 
size, $l_s$. From \eqn{ic1}, $l_s$ is of the order of $ 10^{-29} cm$.  
  
\noindent{\it Phase-2 (Slow-roll inflation) } 

This phase has the typical characteristic that H varies 
very slowly ($\dot H\simeq 0$). 
This corresponds to the plateau region in fig.\ref{fig5}. At the starting
point $t_i\sim 10^4 M_p^{-1}$ the 
value of Hubble parameter is $10^{-4} M_p$. It is clear that 
inflation ends at the time $t_f\sim 5\times 10^5 M_p^{-1}$ where $H\sim 
10^{-5} 
M_p$. During this period the number of e-folds $N_e\simeq 50$ which 
is certainly less than the 
usual  value of 60 e-folds. Thus the inflation ends early when 
universe is just about $10^{-37} sec$ young. 
During this tiny interval scales grow by a factor of $e^{N_e}\simeq 
10^{22}$, so the universe will grow into 
a size as big as $ 10^{-7} cm$ 
starting from the size $10^{-28} cm$ at the start of inflation. 

\noindent{\it Phase-3 (Matter dominated phase)}

At the end of inflation, the matter content starts dominating the 
expansion and the 
Hubble parameter 
then evolves as $H(t)\propto {2\over 3}({1\over t})$ \cite{hs5}. This 
phase 
will last until  $H\approx\sqrt{\Lambda/ 3}$, where $\Lambda$ we would 
like to 
take 
as  
$10^{-122} M_p^2$, the observed value in our universe.
From the simple relation 
$${H(t_1)\over H(t_2)}={t_2\over t_1}$$ 
valid during this expansion we 
can estimate, taking
$t_1\simeq 10^{-37} sec$, $H(t_1)\simeq 10^{-5} 
M_p$, and $H(t_2)\sim 10^{-61} M_p$,  that the age of 
universe is approximately $10^{11} yr$ ($1~yr=3.15\times 10^7 sec$).  Thus 
it gives more or less  the correct age of our universe. 
Also for  most of the time the universe has been dominated by 
matter phase.
Using the scale factor growth as $a(t)\propto t^{2/3}$,
we can estimate that size of the  universe  grows by a factor of 
$10^{38}$ during the matter 
phase. 
Accordingly 
the  present size of our universe is calculated to be $\sim 10^{31} cm$, 
which is 
slightly larger than the present estimates on the size of the 
visible universe  being 
$10^{29} cm$ \cite{wmap,hartle}. 
      
\noindent {\it Phase-4 (de Sitter)}

This is an expected end phase during which universe will
keep on expanding forever like a de Sitter space with constant 
$H=\sqrt{\Lambda/ 3}\sim 
10^{-61} M_p$. This is one of the vacuum solution in the time 
(a)symmetric model. 

\begin{figure}[!ht] \leavevmode
\begin{center}
\epsfysize=5cm
\epsfbox{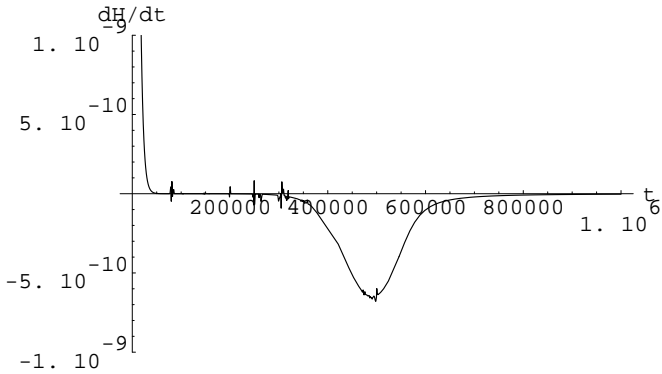}
\end{center}
\caption{ \it The 
plot is for $\dot H$ 
with time-reversal symmetric initial values.  } \label{fig6}
\end{figure}

Let us now plot the rate of change of Hubble parameter during the 
evolution of the universe. The graph in fig.\ref{fig6} corresponds to the
initial values as in fig.\ref{fig5}. The behaviour of expansion resembles 
to our
predictions in section-2. Initially $\dot H>0$, but then it starts to 
slow down and reaches the inflationary phase where $\dot H\sim 0$. It then
further slows down and enters the region which is characterised by $\dot 
H<0$. The evolution then reaches close to the {\it cusp} region near 
about $t=500000/M_p$ where  $\ddot H=0$  briefly. This cusp 
region may be interesting for phenomenology, e.g. reheating of the 
universe etc. Beyond the cusp region $\dot H\propto -{1\over t^2}$, which 
corresponds to the matter dominated era.   

In the end let us comment on the time-reversal asymmetric solution of the 
equations of motion. 
Looking at the 
figures (\ref{fig2}) and (\ref{fig3}) we find that except  near 
$t\simeq0$ the
Hubble parameter evolves more or less in the similar fashion in the two 
models. But there 
will be no phase like $\dot H>0$ for the 
time-asymmetric 
case.  A viable 
inflationary model with time-reversal asymmetric conditions  
will thus be as appealing as the one presented in this work and we will 
not attempt it separately.

 \section{Slow-roll parameters and density perturbations} 

Though it is quite evident from the previous analysis that there is 
genuine slow-roll inflation in the model. Let us find out  the values of 
conventional slow-roll parameters in cosmology. The nearly scale invariant 
spectrum of the large scale 
structure in the universe provides two important conditions on the 
slow-roll parameters 
\bea\lll{slow}
&&\epsilon\equiv{M_p^2\over 2\alpha'} {V'^2\over V^3} \ll 1,~~ 
\eta\equiv{M_p^2\over \alpha'}\left( {V''\over V^2}-{1\ov2}{V'^2\over 
V^3}\right)\ll 1
\eea
where  derivatives are with 
respect to $T$.\footnote{In the conventional form, for the Lagrangian 
with inflaton terms as
$-{1\over 2} \partial_\mu\phi\partial^\mu\phi-V(\phi)$,  the slow-roll 
parameters are given by 
$\epsilon\equiv{M_p^2\over2} {V'^2\over V^2},~ 
\eta\equiv{M_p^2} {V''\over V}$. When compared to the tachyon field in 
our analysis a nontrivial field redefinition $\sqrt{\alpha'V(T)}\,\partial 
T\equiv\partial\phi$ is required which gives the eqs.\eqn{slow}, see also   
\cite{sinha}. } 

Let us evaluate these quantities
for $V(T)=V_0/{\rm cosh}({T\over\sqrt{2}})$, we get 
\bea\lll{slow1}
&&\epsilon={M_s^2\over 8\pi G V_0} {{\rm tanh}^2({T\over\sqrt{2}}){\rm 
cosh}({T\over\sqrt{2}}) \over 4} \br
&&\eta={M_s^2\over 8\pi G V_0} {({3\over2}{\rm 
tanh}^2({T\over\sqrt{2}}) -1){\rm 
cosh}({T\over\sqrt{2}}) \over 2} \ .
\eea
Thus the ratio ${M_s^2\over 8\pi G V_0}$ is an important quantity in 
deciding the slow-roll. If we take 
\be 
{M_s^2\over 8\pi G V_0}\sim O(1) \ ,\lll{slow2}
\ee 
that is what we have 
taken in the numerical analysis, see eq.\eqn{ic1}, then during the plateau 
region where
$T \ll \sqrt{2} $, it provides reasonably smaller values for the slow 
roll parameters. Substituting the values from equations \eqn{ic1} and 
\eqn{ic2} into eq.\eqn{slow1}, it gives for $T \ll \sqrt{2} $
\be
 \epsilon\simeq {T^2\over 24} \ll 1,~~~ \eta\simeq-{1\over 6}.
\ee
This value of $\eta$ is comfortably small but   
is far away from $\eta<<1$ which is 
required for the scale invariance of density spectrum.\footnote{ 
Observationally, the 
spectral index of the density fluctuations, $n_s$, is found to be greater 
than $.94$ and therefore the relation $n_s-1\simeq 
2\eta-6\epsilon$ tells us that $\eta \le .03$.} 
\footnote{ I am thankful to D. Ghoshal, S. Trivedi and A. Sen for 
prompting me to check out 
the possible error in the previous version of the paper.} 
We remark here a few points.
\begin{itemize}
\item     
If we try to have $ {M_s^2\over 8\pi G V_0}\gg
1 $, we are in grave danger of making $\eta \sim 1$ and spoiling the
near scale invariance of the density spectrum. 
\item 
On the contrary, if we consider $ {M_s^2\over 8\pi G V_0}\ll
1 $, we end up in the problem of having initial value of Hubble parameter 
much larger than the string scale. This choice will then 
require potential quantum corrections to be considered in any analysis.
\end{itemize}  
As we just saw, taking $ {M_s^2\over 8\pi G V_0}\sim O(1)$ helps in a 
very subtle manner in producing slow-roll inflation. 

The cosmological measurements 
also put an 
upper bound on  the amplitudes of density perturbations during the 
slow-roll inflation. These  
perturbations grow in size during the inflation and during rest of later 
part the expansion. The COBE measurements give a bound on metric
perturbations \cite{llith} 
\be
\delta_H\equiv {\sqrt{\alpha'}\over\pi\sqrt{75}M_p^3}{V^2\over V'} 
{ \sim}2\times 10^{-5} \ee
This bound is 
applicable nearly $60$ e-folds prior to the end of inflation.
We calculate 
\bea\lll{slow5}
\delta_H&=&{\sqrt{2}\over\pi\sqrt{75}}{ 8\pi G V_0 \over M_s^2} {g_s  
\over \sqrt{v_0}} {\rm cosech}({T\over\sqrt{2}}) \br
&=&{3\sqrt{2}\over\pi\sqrt{75}}10^{-4}{\rm cosech}({T\over\sqrt{2}})\br
&\approx&{6\over\pi\sqrt{75}}{10^{-4}\over T} ~~~~~{\rm for }~T\ll\sqrt{2}
\eea
where we have used $M_s={g_s \over\sqrt{ v_0}}M_p=10^{-4}M_p$ and the 
exact values in \eqn{ic1}. Also note that from \eqn{ic2} the smallest 
value of $T$ is $T_0=10^{-3}$ in this model.

Here, it is 
significant to  note that our chosen value 
$H \simeq M_s=10^{-4} M_p$ at the start of inflation is almost  
right.  To see this we need to know 
the value of $T(t)$ at the time when fluctuations seed in. 
From the fig.\ref{fig5}, we only have about $50$ 
e-foldings in total, hence the
quantum fluctuations must seed in at the start of the 
inflation itself in order to comply with the COBE normalisations. 
At the beginning of the fluctuations it is reasonable to take
$T\sim 10^{-2}$. From \eqn{slow5} this gives us  
$$\delta_H\sim 10^{-3}$$ 
at the beginning of the last 50 e-folds. So there is 
slight problem with the COBE limits. The amplitudes we get are 
larger but those are not too
 far away from the observational limits.  
  
 We note that there is  enough 
flexibility in the model in order to get $\delta_H$ right. For example, 
looking at \eqn{slow5} we find that a value of ${g_s\over 
\sqrt{v_0}}$ less than $10^{-4}$ 
will be more favourable. It means, the string scale should be taken 
further 
small than $10^{-4} M_p$. Alternatively,  the ratio ${ 8\pi G V_0 
\over M_s^2}$ which we have taken $\sim 1$  can be relaxed a bit and can 
be taken slightly greater than 1.
That is, $H_m/M_s$ should be slightly greater than 1. This will allow 
a larger initial value of $T(0)$, while maintaining the same number of 
e-folds $N_e\sim50$. However, $H_m>M_s$  might invite quantum 
stringy corrections as we stated before. Still a purely classical 
analysis may provide some useful insight.   
On the opposite side, from 
eq.\eqn{slow5} it will also be prudent to choose $ 8\pi G V_0$ smaller 
than $M_s^2$, but it should be done only marginally, otherwise a large 
difference will
interfere with slow-roll limits on $\epsilon$ and $\eta$. Particularly 
$\eta$ 
is very much sensitive to this. This last 
point corresponds to the choice
in \eqn{sit2}, which incorporates a critical number density of the branes.
However, this later choice does not seem to be a better choice as it also 
has the potential to reduce the number 
of e-folds further down from 50. Amongst all of these, $H_m$ being 
slightly greater 
than $M_s$ is a better choice  accompanied with a larger initial value of 
$T(0)$. 

Thus any possible 
remedy of the large values of $\eta$ and $\delta_H$ has to necessarily involve 
taking ${g_s\over\sqrt{v_0}}<10^{-4}$ and $H_m$ slightly greater than 
$M_s$ but {\it not like $H_m\gg M_s$}.
In the next  we demonstrate rolling solutions which include both 
of these conditions.

\vskip.5cm
\noindent\underline{\bf Improved results with 
${g_s\over\sqrt{v_0}}<10^{-4}$ and 
$H_m>\sim M_s$ :}
We basically present two examples which incorporate 
${g_s\over\sqrt{v_0}}<10^{-4}$ 
and $H_m> M_s$.

\begin{figure}[!ht]
\leavevmode
\begin{center}
\epsfysize=5cm
\epsfbox{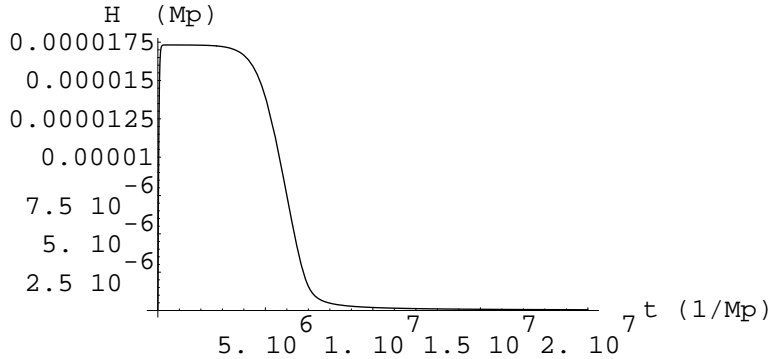}
\end{center}
\caption{ \it The 
plot is for $H(t)$ 
with time-reversal symmetric initial values 
$H(0)=0,~ T(0)=10^{-2}$ with ${8\pi G 
V(0) \over 3}=3\times 10^{-10}  $. 
The value of  $N_e\sim 85$. } \label{fig7}
\end{figure}

\noindent{\it Example-1}: Let us consider 
taking the parameters as
\be\label{ic1q}
{g_s\over\sqrt{v_0}}=10^{-5}, ~M_s=10^{-5}M_p,~~ 8\pi G V_0=9\times 
10^{-10} M_p^2 
\ee
with the time-symmetric initial conditions  at $t=0$ as
\be\label{ic2q}
T(0)=10^{-2},~ \dot a(0)=0, ~a(0)=\sqrt{3/ (8\pi G V_0)} \ .
\ee
That is, the string scale is $M_s=10^{14} GeV$ and $H_m=\sqrt{3}M_s$.
Note we can still trust a purely classical analysis.   
We have plotted the result in the  figure (\ref{fig7}) for the rolling
solution of eqs.\eqn{aeqn1} with the above initial conditions.

With this parametric choice we find that  $\eta \simeq 
-{1\over 18}$, which gives $n_s=0.89$ and thus is much suitable for 
slow-roll inflation compared to the value of $\eta \simeq 
-{1\over 6}$ and may 
produce scale-invariant density spectrum. 
For the size of the amplitudes at
60 e-folds prior to the end of inflation, we now find that 
$$\delta_H|_{60\,e-folds} \simeq 1.3\times 10^{-4},$$ which is only 
marginally larger than the cosmological limits. This shows that there
is definite improvement in the over-all picture. 

\noindent{\it Example-2}: Let us now consider 
another set of values
\be\label{ic1qa}
{g_s\over\sqrt{v_0}}=10^{-5}, ~M_s=10^{-5}M_p,~~ 8\pi G V_0=12\times 
10^{-10} M_p^2 
\ee
with the same initial conditions  as in \eqn{ic2q}.
The string scale is still $M_s=10^{14} GeV$ but
$H_m={2}M_s$. 
We find that this model gives $\eta\sim - {1\over 24}$ and the amplitudes 
as    
$$\delta_H|_{60\,e-folds} \simeq 2.6\times 10^{-5},$$ 
which is what one requires for the phenomenology.

Thus, essentially we have discovered that  
a string mass scale as low as $10^{14} GeV$ 
together with a maximum value of Hubble parameter in the range $2 M_s\ge 
H_m\ge \sqrt{3} M_s$ can produce the 
desired slow-roll results within the reach of observational bounds on 
$\epsilon, \eta$ and 
$\delta_H$.

\section{Conclusions}

We have studied rolling tachyon cosmological solutions in a
de Sitter spacetime. The Friedman-Robertson-Walker spacetimes are studied
with both time-reversal symmetric as well as non-symmetric initial 
conditions. As stated by Sen \cite{sen3} the time-symmetric conditions 
are the ones which are favoured by the open-string completeness 
conjecture. Although, the time asymmetric conditions are equally allowed 
but the corresponding open-string picture is not quite understood.
From our study  of the FRW solutions of the equations of motion,
we find that from phenomenological point of view it is hard to distinguish 
between the two type of situations. Particularly, given the tachyon 
potential, the evolution of the 
Hubble parameter is strikingly similar except near $t=0$, in both the 
situations. The 
configurations are completely non-singular in both the cases. 

In order to get a viable cosmological models of inflation, we need to have
at least $8\pi G V_0 \simeq M_s^2$  with 
$M_s\simeq 10^{15} GeV$ or lower. With large volume 
compactification ($v_0>1$), this construction heavily depends upon a 
large number density of unstable 3-branes over the six-volume of the 
compact manifold. This is because we 
must have  ${N g_s\over v_0}\sim O(1)$ in order that $H_m\sim M_s$, 
which helps in achieving 
inflation.
We define a {\it critical number 
density} for which there is single 3-brane per unit (string-size) 
six-volume  of the compact manifold.  

We find that near $t=0$ the universe will have an  energy 
density of the order of $10^{78} gm.cm^{-3}$ and not the Planck density  
$10^{94} gm.cm^{-3}$ as in big-bang scenario. The reason is that 
we cannot afford to have Planck density in these models at $t=0$, if we 
want appropriate amplitudes for the density fluctuations during 
the inflation. In the section-3,  
we have analysed a particular model in detail
where solutions give rise to roughly 50 e-folds of inflation. 
Due 
to the restricted number of e-folds in this model, the 
fluctuations which give rise to the density perturbations, must set in at 
the very beginning of the inflation. Note that, during 
inflation the energy density of universe remains {\it almost} fixed. This 
is because the inflation is largely driven by the initial de Sitter 
like phase. Therefore at the  time $t\simeq 10^{-37} sec$ when inflation 
ends we will measure $H\sim 10^{14} GeV$
with an energy density approximately $10^{78} gm.cm^{-3}$. These are more 
or less the standard values  at the end of inflation in standard big-bang 
model. After the inflation the matter dominated phase starts 
which lasts until when $H\approx 10^{-61} M_p$. 

 We also find that
during the large part of its evolution the universe has remained matter 
dominated. 
The calculated {\it age of universe} comes out to 
be about $\sim 10^{11} yr$ which is quite admissible. Also the {\it size 
of visible 
universe} comes out to be $\sim 10^{30} cm$, which is just right in 
magnitude when compared to cosmological data.    
 What about the radiation dominated phase? The answer is; we do not 
explicitly encounter a radiation dominated 
era where scale factor would grow as $a(t)\propto t^{1\over2}$. But since 
the expansion switches in time
from $\dot H >0$ (early de Sitter phase) to  $\dot 
H\propto -{2\over3}t^{-2}$ 
(matter dominated phase) through an intermediate value $\dot H\simeq 0$ 
(inflationary 
phase), somewhere in between (just after the inflation) the universe may
pass through radiation dominated era.

In summary, it seems that inflationary tachyon models of the type 
presented in this work may be
suitable for cosmological applications. 
We have found from our analysis that having
string mass scale as low as $10^{14} GeV $ 
together with $2 M_s\ge H_m\ge \sqrt{3} M_s$ does produce the 
desired slow-roll results within the  observational limits on 
$\epsilon, \eta$ and $\delta_H$.   
However, there are limitations. 
These models require large 
number density of the non-BPS branes. Whether it can be admitted 
 is an open question.    
From string theory point 
of view such a construction
does provide a viable alternative to those models which assume
spacetime warping \cite{sinha}. 
In cosmology, particularly, we have to ask an important question; Is 
it that our 
universe never had a Planck density phase? Can we do without Planck 
density phase at the big-bang in cosmology? The scenarios we discussed in 
this paper
do not support a Planck density phase. However, we 
do not claim to know the complete answer.

\leftline{\bf Acknowledgment:}  

I would like to thank  Shibaji Roy for careful
reading of the manuscript and for the useful discussion that followed.
I am extremely thankful to the referee for the comments which have 
resulted in the major improvements of this manuscript.

\end{document}